\documentclass[pre,twocolumn,showpacs,superscriptaddress]{revtex4}
\usepackage{amsmath,amssymb}
\usepackage[pdftex]{graphicx}

\begin{document}
\title{%Compared 
Comparative quantum and semi-classical analysis of Atom-Field Systems I: density of  states and excited-state quantum phase transitions}
\author{M. A. Bastarrachea-Magnani}
\affiliation{Instituto de Ciencias Nucleares,
Universidad Nacional Aut\'onoma de M\'exico, Apdo. Postal 70-543, Mexico D. F., C.P. 04510}
\author{S. Lerma-Hern\'andez}
\affiliation{Departamento de F\'isica, Universidad Veracruzana, Circuito Aguirre Beltr\'an s/n, Xalapa, Veracruz, M\'exico, C.P. 91000}
\email{slerma@uv.mx}
\author{J. G. Hirsch}
\affiliation{Instituto de Ciencias Nucleares,
Universidad Nacional Aut\'onoma de M\'exico, Apdo. Postal 70-543, Mexico D. F., C.P. 04510} 

\date{today}
\begin{abstract}
We study the non-integrable Dicke model and its integrable approximation, the Tavis-Cummings model,
as functions of both the coupling constant and the excitation energy.
Excited-state quantum phase transitions (ESQPT) are 
found analyzing 
the density of states in the semi-classical limit and comparing it with  numerical results for the quantum case in large Hilbert spaces, taking advantage of  efficient methods recently developed. Two different  ESQPTs are identified  in both models,  which are signaled  as   singularities in the semi-classical density of states, one {\em static} ESQPT occurs for any coupling, whereas a {\em dynamic} ESQPT is 
observed
only in the superradiant phase. The role of the unstable fixed points of the Hamiltonian semi-classical flux in the occurrence of the ESQPTs is discussed and determined. Numerical evidence is provided that shows that  the   semi-classical results describe very well the tendency of the quantum energy spectrum for any coupling in both models. Therefore the semi-classical density of states can be used to study the statistical properties of the fluctuation in the spectra, a study that is presented in 
a companion paper. 
\end{abstract}
\maketitle

%%%%%%%%%%%%%%%%

%%%%%%%%%%%%%%%%
\noindent

%%%%%%%%%%%%%%%%

\section{Introduction}
The Dicke Hamlltonian describes a system of $\mathcal{N}$ two-level atoms interacting with a single monochromatic electromagnetic radiation mode within a cavity \cite{Dicke54}. In the language of quantum computation, it can also describe a set of $\mathcal{N}$ qubits from quantum dots, Bose-Einstein condensates or QED circuits \cite{Sche03,Sche07,Blais04,Fink09}, interacting through a bosonic field. The Hamiltonian is very simple but not exactly solvable, and continues to drive research into its properties. The most representative feature of the Dicke Hamiltonian is its second-order quantum phase transition (QPT) in the thermodynamic limit \cite{Hepp73,Wang73}. The ground state of the system goes from a normal to a superradiant state when the atom-field interaction reaches a critical value. This transition is an example of a quantum collective behavior \cite{Nah13}. 
The interest on solving the Dicke Hamiltonian for a finite $\mathcal{N}$ comes not only from the fact that it provides a good description for the systems manipulated in the laboratory, but from the close connection found between entanglement, quantum phase transitions, and quantum chaos \cite{Emary03,Lam05,Vid06}. Recently  Dicke-like Hamiltonians have attracted much attention because of the experimental realization of the superradiant phase transition in a BEC \cite{Bau10,Nag10}, while the debate around the validity of the description and its relation with the no-go theorem is far from closed \cite{Kni78,Bial79,Gaw81,Lib04,Nataf10,Vieh11,Cuiti12}.   
In the thermodynamic limit (equivalent in the present models to the semi-classical limit), when the  number of atoms $\mathcal{N}$ goes to infinity, the mean field description becomes exact, and a Holstein-Primakoff expansion around it provides analytic solutions \cite{Emary03}, which allow to extract the critical exponents for the ground state energy per particle, the fraction of excited atoms, the number of photons per atom, their fluctuations and the concurrence \cite{Emary03,Lam05,Vid06,Chen0809}. For a finite number of atoms $\mathcal{N}$, the model is in general non-integrable, and care must be taken when the first order in the $1/\mathcal{N}$ expansion is employed because of its singular behavior around the phase transition \cite{OCasta11a,OCasta11,Hir13}. 

The existence of an excited-state quantum phase transition (ESQPT) in the Dicke and Tavis-Cummings (TC) models was recently pointed out by Perez-Fern\'andez, et. al. \cite{Per11A}. An ESQPT takes place along the energy spectrum, for fixed values of the Hamiltonian parameters. It is manifested by singularities in the level density, order parameters, and wave function properties \cite{Cap08}. The ESQPTs have been analyzed in several nuclear physics  models \cite{Cej06} and could have important effects in decoherence \cite{Rel08} and the temporal evolution of quantum quenches \cite{Per11B}. 
Their relationship with the ground state QPT is not completely clear, so the issue is open to current research.

We consider the non-integrable Dicke model and its integrable approximation, the Tavis-Cummings model, where the counter-rotating terms are neglected. These models are studied as  functions of the coupling between atoms and field and as  functions of the energy. The excited-state quantum phase transitions in these models are identified by studying the density of states in the semi-classical limit. We identify two ESQPT of different nature, one {\it static} appearing for any coupling and a {\em dynamic} ESQPT which is present only in the superradiant phase. The role of the unstable fixed points, where abrupt changes in the available phase take place, in determining the ESQPTs is exposed.  Analytic expressions for the density of states are obtained which coincide with those  derived by T. Brandes recently \cite{Bran13}.  We compare the semi-classical results  with numerical results of the quantum model in large Hilbert spaces, taking advantage of  efficient methods recently developed \cite{Chen0809,Basta11,Basta12}. The comparison shows that the semi-classical results describe very well the tendency of the quantum spectra, both in the TC and Dicke models and for the normal and superradiant phases. Consequently the semi-classical density of states can be used to study the statistical properties of the quantum spectrum fluctuations, a study that  is presented in the companion paper \cite{Basta2} of this series of two papers, where additionally  the quantum results are compared with the onset of irregular trajectories in the  semi-classical phase space.                
 
The article is organized as follows: in Section II we present the Dicke and the Tavis-Cummings Hamiltonians and summarize some of their properties. The classical  Hamiltonians are described in Section III together with the analysis of the stable and unstable fixed points. In section IV the available phase space volume as a function of coupling and energy is used  to determine the semi-classical density of states. This density is   compared with  the quantum result in the same section. Section V contains the conclusions. 
%%%%%%%%%%%%%%%%

\section{Dicke and Tavis-Cummings Hamiltonians}

The Dicke model describes the interaction between a system of $\mathcal{N}$ two-level atoms and a single mode of a radiation field within a cavity. The Hamiltonian is made of three parts: one associated to the monochromatic quantized radiation field, a second one to the atomic sector, and a last one which describes the interaction between them. The Dicke Hamiltonian can be written as
\begin{equation}
H_{D}=\omega a^{\dagger}a+\omega_{0}J_{z}+\frac{\gamma}{\sqrt{\mathcal{N}}}\left(a+a^{\dagger}\right)\left(J_{+}+J_{-}\right).
\end{equation}
The frequency of the radiation mode is $\omega$, which has an associated
photon
 number operator $a^{\dagger}a$. For the atomic part $\omega_{0}$ is the excitation energy, meanwhile $J_{z}$, $J_{+}$, $J_{-}$, are collective atomic pseudo-spin operators which obey the SU(2) algebra. It holds that if $j(j+1)$ is the eigenvalue of $\mathbf{J}^{2}=J_{x}^{2}+J_{y}^{2}+J_{z}^{2}$, then $j=\mathcal{N}/2$ (the pseudo-spin lenght) defines the symmetric atomic subspace which includes the ground state. $\gamma$ is the interaction parameter. For atomic systems, it depends principally on the atomic dipolar moment. Besides, $H_D$ commutes with the parity operator $\Pi$,
\begin{equation} \label{parity}
\Pi=e^{i\pi\Lambda},\,\,\,\mbox{with}\,\,\,\Lambda=a^{\dagger}a+J_{z}+j.
\end{equation}
The eigenvalues of the $\Lambda$ operator, $\lambda=n+m+j$ , are the total number of excitations, where $n$ is the number of photons and $n_{exc}=m+j$ the number of excited atoms. As it was mentioned, in the thermodynamic limit a second-order QPT takes place when the interaction parameters reaches the critical value $\gamma_{c}=\sqrt{\omega\omega_{0}}/2$, separating the system in two regions, the normal phase ($\gamma<\gamma_{c}$) and the superradiant phase ($\gamma>\gamma_{c}$). 
In the normal phase the ground state has $\lambda = 0$, {\em i.e.} no photons and all atoms in their ground state.
The superradiant phase is characterized by a macroscopic population of the upper atomic level
and a comparable average photon number in the ground state of the system.  

In general, for finite $\mathcal{N}$ the Dicke Hamiltonian is not integrable. However, it has two integrable limits: when $\gamma\rightarrow 0$ and when $\omega_o\rightarrow 0$ \cite{Basta11}. Moreover, when the coupling is weak it is possible to make the {\it Rotating Wave Approximation}, by ignoring the counter-rotating terms. The result is another integrable limit, the Tavis-Cummings Hamiltonian \cite{TC68}
\begin{equation}
H_{TC}=\omega a^{\dagger}a+\omega_{o}J_{z}+\frac{\gamma}{\sqrt{\mathcal{N}}}\left(aJ_{+}+a^{\dagger}J_{-}\right).
\end{equation}
The TC Hamiltonian is integrable because it commutes with the $\Lambda$ operator. Its  conserved eigenvalues  $\lambda$ define a set of subspaces where $H_{TC}$ can be diagonalized independently. It also has a QPT in the thermodynamical limit, when the coupling has a critical value of $\gamma_{c,TC}=\sqrt{\omega_{o}\omega}$. For couplings $\gamma \leq \gamma_{c, TC}$, the ground state is the state with $\lambda = 0$, with  no photons nor excited atoms, as in the Dicke model. 
When  $\gamma > \gamma_{c, TC}$ the ground state has a certain $\lambda_c>0$, which grows monotonically with $\gamma$. As an integrable approximation of the Dicke model, the TC model will help us to gain understanding of the connection between chaos, integrability and the ESQPT. 

We can write both models in one expression,
\begin{equation}
\begin{split}
H&=\omega a^{\dagger}a+\omega_{0}J_{z}+\\
&+\frac{\gamma}{\sqrt{\mathcal{N}}}\left[ \left(aJ_{+}+a^{\dagger}J_{-}\right) +\delta\left(a^\dagger J_{+}+a J_{-}\right)\right],
\end{split}
\end{equation}
where $\delta=0$ and $1$ for the TC and Dicke models, respectively. With this parametrization  the QPT's critical values are  $\gamma_{c}=\sqrt{\omega_{0}\omega}/(1+\delta)$.
From now on, we will focus on the subspace with largest pseudo-spin, where $j=\mathcal{N}/2$.
%%%%%%%%%%%%%%%%

\section{Classical Hamiltonians}
As it has been discussed in previous works for the Dicke \cite{DickeBrasil,OCasta11a,OCasta11,Hir13,Bran13} and Tavis-Cummings \cite{OCasta09} models, many insights can be gained by studying  the classical limit. Since we chose $\hbar=1$, this limit is equal to the thermodynamical limit $j\rightarrow \infty$.  

The classical versions of the Dicke and TC  models can be obtained employing the {\it naive} substitution of the pseudospin variables by classical angular momentum ones ($J_i\rightarrow j_i$), and the substitution of the boson variables by a classical harmonic oscillator with $m\omega=1$ ($\sqrt{2}a\rightarrow q+ip$ and $\sqrt{2}a^\dagger\rightarrow q-ip$). Recalling the  relations $J_+=J_x+i J_y$ an $J_-=J_x-i Jy$, we obtain
\begin{equation}
H_{cl}= \omega_o j_z+\frac{\omega}{2}(q^2+p^2)+ \frac{\gamma}{\sqrt{j}}\left[(1+\delta) q \, j_x -(1-\delta) p \, j_y\right].
\end{equation}
In reference \cite{DickeBrasil} it was shown that the previous  Hamiltonian is entirely equivalent to that obtained by  using bosonic and $SU(2)$ coherent states. The pseudospin  variables satisfy  the  Poisson-bracket algebra $\{j_i,j_j\}=\epsilon_{ijk} j_k$. 
%From there c
Canonical variables satisfying $\{P,Q\}=-1$ can be 
%obtained 
constructed from them
as $P=j_z$ and $Q=\phi=\tan^{-1}
%(S_y/S_z)
(j_y/j_x)$, 
where $\phi$ is the azimuthal angle of the vector $\vec j=(j_x,j_y,j_z)$ whose magnitude is constant $|\vec j|=j$. In terms of the canonical variables the classical Dicke and TC Hamiltonian reads 
\begin{eqnarray}
H_{cl}&=& \omega_o \, j_z+\frac{\omega}{2}(q^2+p^2)+ \label{DiHam}\\
      &  &\gamma \sqrt{j} \sqrt{1-\frac{j_z^2}{j^2}}\left[ (1+\delta) \, q \cos\phi -(1-\delta) \, p\sin\phi\right].\nonumber
\end{eqnarray} 
%From here the 
The associated classical
equations of motion are 
\begin{eqnarray}
\frac{dq}{dt}&=&\frac{\partial H_{cl}}{\partial p}= \omega \, p- (1-\delta)\gamma \sqrt{j} \sqrt{1-\frac{j_z^2}{j^2}}  \sin\phi \label{qp}\\
\frac{dp}{dt}&=&-\frac{\partial H_{cl}}{\partial q}= -\omega \, q-(1+\delta)\gamma\sqrt{j} \sqrt{1-\frac{j_z^2}{j^2}}\cos\phi  \label{pp}\\
\frac{d\phi}{dt}&=&\frac{\partial H_{cl}}{\partial j_z}=\omega_o  \\
& -&\frac{\gamma j_z }{j^{3/2} \sqrt{1-\frac{j_z^2}{j^2}}} \left[ (1+\delta) q \cos\phi -(1-\delta) p\sin\phi\right] \label{php}\nonumber\\
\frac{d j_z}{dt}&=&-\frac{\partial H_{cl}}{\partial \phi}=2\gamma\sqrt{j}\sqrt{1-\frac{j_z^2}{j^2}}\nonumber\\
 &\times& \left[(1+\delta) q\sin\phi +(1-\delta) p\cos\phi\right]\label{jp}
\end{eqnarray}

The fixed points of the Hamiltonian flux correspond to the values $(q_m,p_m,j_{zm})$ which produce the simultaneous cancellation of the four derivatives.  
Two of them are present for any value of the coupling constant $\gamma$,  
 $$(q_m,p_m,j_{zm})=(0,0,\pm j).$$ 
Note that $j_z=\pm j$ correspond to the north and south pole of the pseudospin  sphere where the value of the azimuthal angle is irrelevant. If we evaluate the Hamiltonian in the previous fixed points, we obtain, respectively and for any coupling, the energies $\epsilon=\pm 1$; where we have, conveniently, rescaled the energy as
\begin{equation}
\epsilon\equiv \frac{E}{\omega_o j}.
\end{equation}
The nature of the previous fixed points is as follows, the point $(q_m,p_m,j_{zm})=(0,0,+j)$ is an unstable fixed point for any value of the coupling $\gamma$, whereas the point $(q_m,p_m,j_{zm})=(0,0,-j)$ is a stable fixed point for couplings  $\gamma\leq\gamma_c$ that  becomes  unstable for couplings $\gamma>\gamma_c$. 
It represents the semiclassical description of the ground state in the normal phase, with no photons and no excited states.

 For couplings larger than the critical one, new stable points emerge whose properties depend on the model we are considering, Dicke ($\delta=1$) or TC ($\delta=0$). For the Dicke model two degenerate stable fixed points emerge which are  given by
\begin{eqnarray}
(q_m,p_m)_\pm &=& \left(\mp\frac{2\gamma\sqrt{j}}{\omega} \sqrt{1-\left(\frac{\gamma_c}{\gamma}\right)^4}, 0\right)\\
(\cos\phi_m, j_{zm})_\pm &=& \left(\pm 1, -j\left(\frac{\gamma_c}{\gamma}\right)^2 \right),\label{minSP}
\nonumber
\end{eqnarray}
whereas for the integrable  TC model a continuous set of stable fixed points parametrized by the angle $\phi\in[0,2 \pi)$ appear which are given by 
\begin{eqnarray}
(q_m,p_m)&=&  \frac{\gamma\sqrt{j}} {\omega} \sqrt{1-\left(\frac{\gamma_{c}}{\gamma}\right)^{4}}\left(-\cos\phi, \sin\phi \right)\nonumber\\
j_{zm}&=& -j\left(\frac{\gamma_c}{\gamma}\right)^2.
\end{eqnarray} 
The continuous set of fixed point in the TC model is consequence of the symmetry associated with the conserved quantity $\Lambda$ (\ref{parity}), whose classical version is $\Lambda_c=(q^2+p^2)/2+j_z+j$.  

To better visualize the properties of the fixed points in both models, we construct  energy surfaces  in terms of the pseudo-spin variables $j_z$ and $\phi$.
%  to this end we 
Equating to zero Eqs.  (\ref{qp}) and (\ref{pp}) we obtain $\sqrt{j}\omega p=(1-\delta)\gamma \sqrt{j^2-j_z^2}\sin\phi $ and $\sqrt{j}\omega q=-(1+\delta)\gamma\sqrt{j^2-j_z^2} \cos\phi$, by substituting these results in the Hamiltonian we obtain 
a semiclassical expression for the energy as a function of  $j_z$ and $\phi$,
\begin{equation}
\frac{E (j_z,\phi)}{\omega_o j}=\frac{j_z}{j}-\frac{\gamma^2}{2\gamma_c^2}\left(1-\frac{j_z^{2}}{j^{2}}\right)\left(1-\frac{4\delta}{(1+\delta)^{2}} \sin^2\phi\right).
\label{ensur}
\end{equation}
\begin{figure}
\begin{tabular}{ccc}
\qquad  $\gamma=0.2 \gamma_c$& $\gamma=1.0 \gamma_c$& $\gamma=2.0 \gamma_c$\\
\rotatebox{90}{\ \ \ \ \ \   \Large{$\theta\sin\phi$}}\includegraphics[angle=0,width=0.145\textwidth]{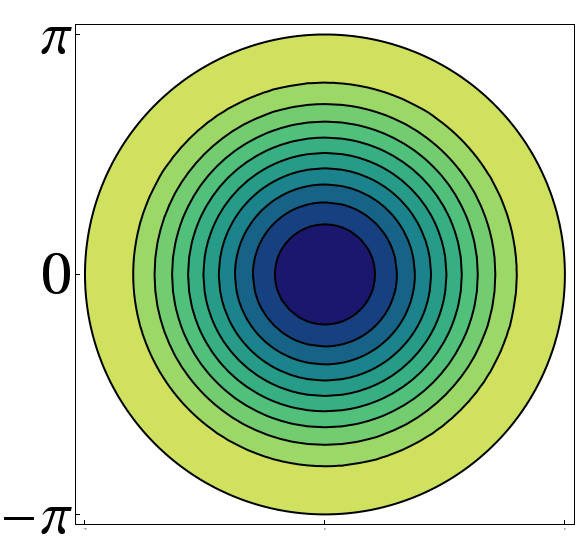}
&\includegraphics[angle=0,width=0.132\textwidth]{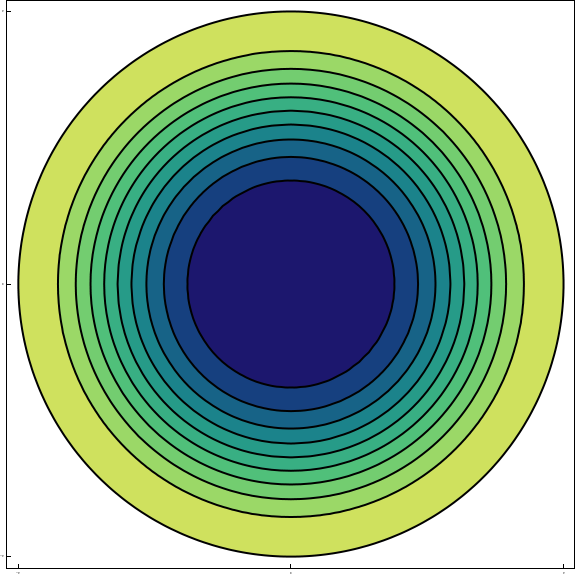}
&\includegraphics[angle=0,width=0.132\textwidth]{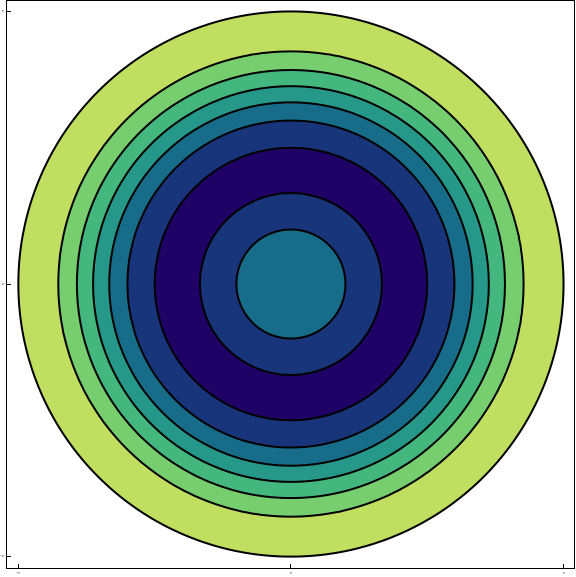}
\\
\rotatebox{90}{\ \ \ \ \ \ \ \ \    \Large{$\theta\sin\phi$}}\includegraphics[angle=0,width=0.15\textwidth]{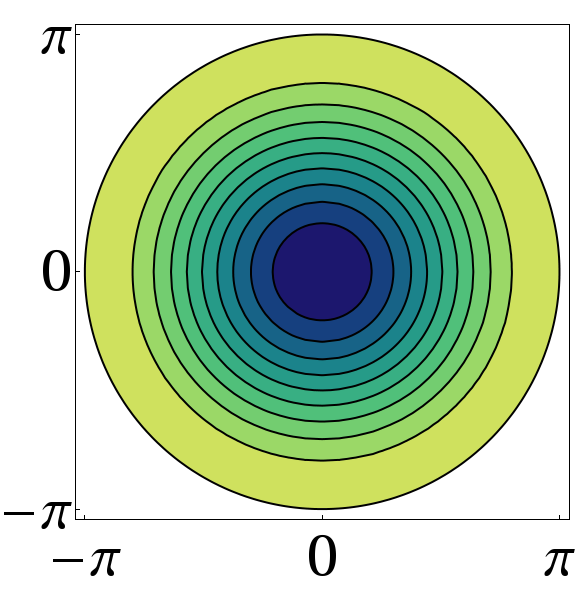}
&\includegraphics[angle=0,width=0.138\textwidth]{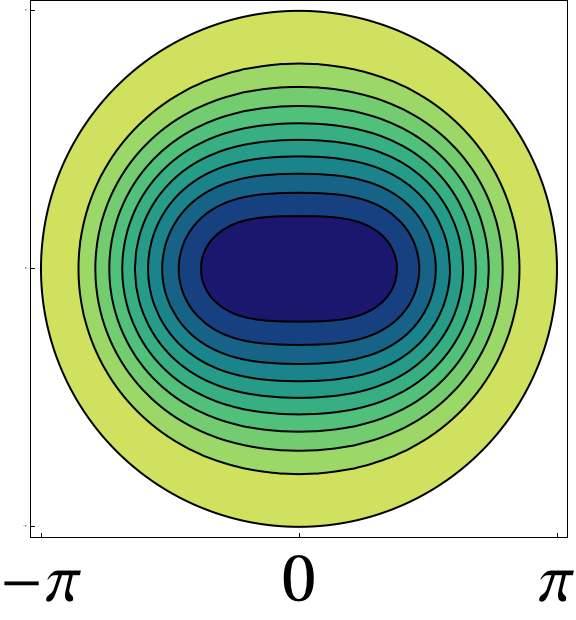}
&\includegraphics[angle=0,width=0.138\textwidth]{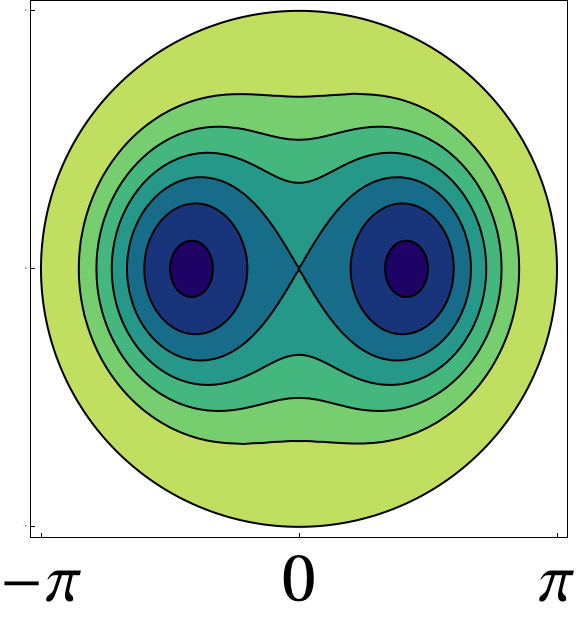}
\\   & \Large{$\theta\cos\phi$} & 
\end{tabular}
\caption{(Color online) Contour plots  of the energy surface, Eq.(\ref{ensur}), for the Tavis-Cummings (above) and Dicke (below) models, for three different couplings. Dark tones indicate low values of the energy. Angular variables of the pseudo-spin $\vec j$ are used: $\phi$ is the azimuthal angle and $\theta$ is  the zenith angle measured respect to the south pole ($j_z=-j\cos\theta$).}  
\label{FigEneSur}
\end{figure} 
For the TC model ($\delta=0$) the
% previous 
energy surface is independent on the angle $\phi$. In Fig.\ref{FigEneSur}  contour plots of the 
%previous 
energy surface are shown for the TC and Dicke models and  for three different 
values of the
couplings. Variables $\phi$ and $\theta$ are used, where $\theta$ is the zenith angle of $\vec{j}$ measured respect to the south pole ($j_z=-j \cos\theta$).  Because the  symmetry $\Lambda$ of the  TC model,  its contours are circular for any coupling.  For small couplings the energy surfaces of the Dicke and TC models are almost indistinguishable, with circular contours and a global minimum in the south pole ($\theta=0$).   For couplings close but below the critical value, the global  minimum is yet  the south pole but the contours  begin to exhibit deformation
in the Dicke model. For couplings above $\gamma_c$, the south pole becomes a local maximum for the TC model and a saddle point in the Dicke model. Besides, according to Eq. (\ref{minSP}), two degenerate minima appear in the case of the Dicke model in  $\phi=0$ and $\pi$,  whereas for the TC model,  the energy surface  takes a mexican hat form with a continuous set of minima circularly located around the south pole, which is related to a Goldstone mode \cite{Baksic13,Xiang13}.      
      
\begin{figure}
\begin{tabular}{c}
\rotatebox{90}{\ \ \ \ \ \ \ \ \   \qquad  \qquad \LARGE{ $\mathbf{\epsilon}$} }\includegraphics[angle=0,width=0.45\textwidth]{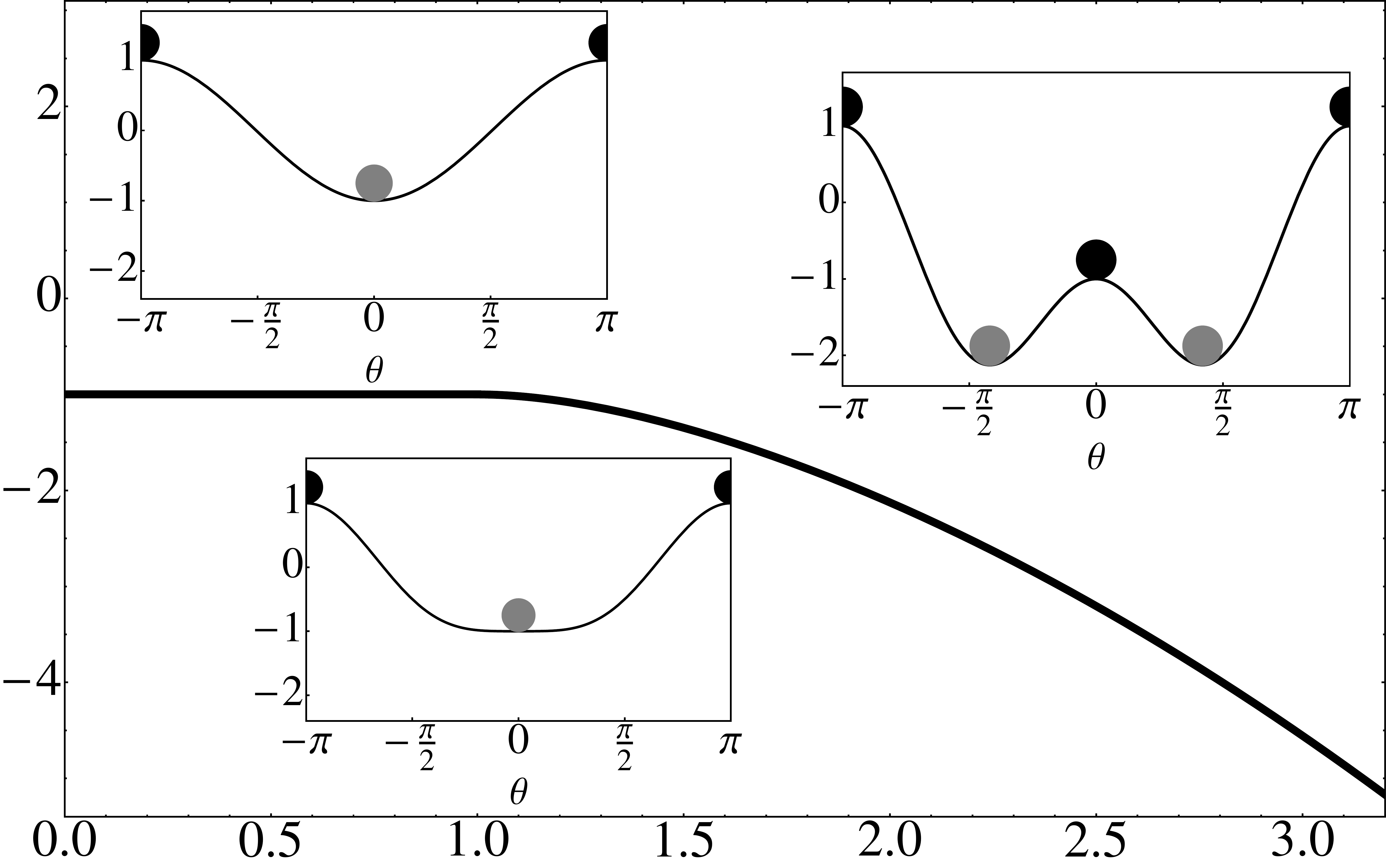}%
\\\Large{ $\mathbf{\gamma/\gamma_c}$}
\end{tabular}
\caption{
 Scaled energy minimum ($\epsilon_{min}\equiv E_{min}/(\omega_o j)$) as a function of the coupling constant measured respect to the critical value ($\gamma/\gamma_c$). In the insets  three typical  energy surfaces are shown for couplings, from left to right,  $\gamma/\gamma_c=0.2$, $1.0$, and $2.0$.   Stable and unstable fixed points are signaled by gray and  black circles, respectively. The angle $\theta$ is that formed by the pseudo-spin $\vec {j} $ and the negative z-axis. }
\label{fig0}
\end{figure}

The  energy  minimum  is  obtained by evaluating the Hamiltonian in the  stable fixed points. The result, valid for both the TC and Dicke models, is given by
\begin{equation}
\epsilon_{min}\equiv\frac{E_{min}}{\omega_o j}=\left\{\begin{array}{lr} -1 & {\hbox{for }} \gamma\leq \gamma_c  \\
-\frac{1}{2} \left(\frac{\gamma_c^2}{\gamma^2}+ \frac{\gamma^2}{\gamma_c^2}\right) &{\hbox{for }} \gamma > \gamma_c\\ \end{array} \right. .
\label{enesClass}
\end{equation}
This function is shown in Fig \ref{fig0}, together with cuts of the energy surface (\ref{ensur}) for $\sin\phi=0$.  The cuts  are shown as a function of the  angle $\theta$, where  positive and negative $\theta$ correspond, respectively,  to $\phi=0$ and $\phi=\pi$. The fixed points and their respective nature can be %clearly 
easily
visualized in these energy surfaces, 
%likewise it is clear 
and it is apparent
that the transition that takes place in the critical coupling is a second order pitchfork transition. 

%%%%%%%%%%%%%%%%
%%%%%%%%%%%%%%%%

\section{Density of states}
The stable fixed points  of the classical TC and Dicke models (gray dots in Fig.\ref{fig0}), identified and discussed in the previous section, are useful  to understand the behavior of the energy minimum, %(or 
associated with the
ground state Quantum  Phase Transition in the quantum version of the models.
Likewise, the unstable ones  (black dots in Fig.\ref{fig0}) 
%allow to identify interesting properties of the models, because they 
are benchmarks in the energy space which indicate abrupt  changes in the behavior of the available phase space. These changes, whose quantum analogues are referred to as  excited-state quantum phase transition  \cite{Cap08},  deserves a detailed analysis which will be conducted in the following.   

\subsection{Classical volume of the available phase space} 

\begin{figure*}
\centering{
\begin{tabular}{ccc}
\vtop{\vskip -0.290\textwidth \hbox{\rotatebox{90}{\  \ \   \quad\quad \quad\quad \quad \Large{$\omega \nu(\epsilon)/(2 j)$}}  \ \ \includegraphics[angle=0,width=.316\textwidth]{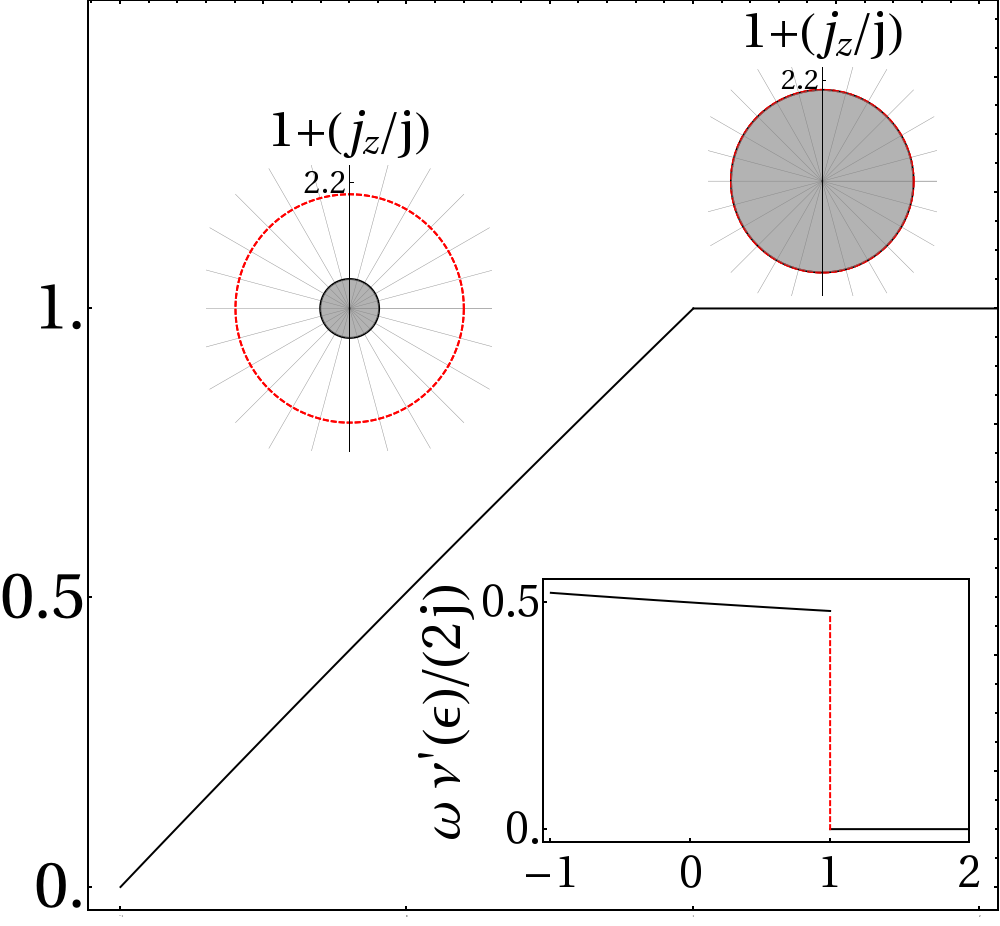}
}}&
\includegraphics[angle=0,width=0.29\textwidth]{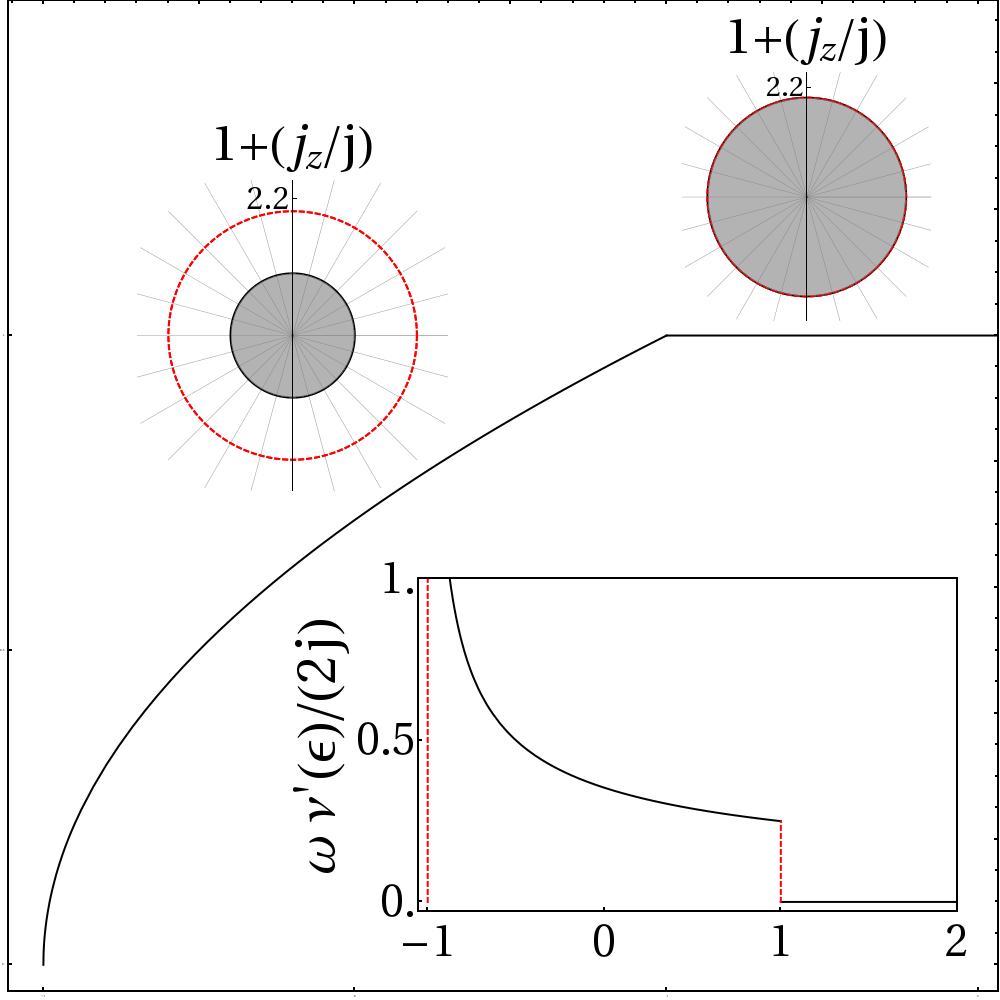}
&
\includegraphics[angle=0,width=0.29\textwidth]{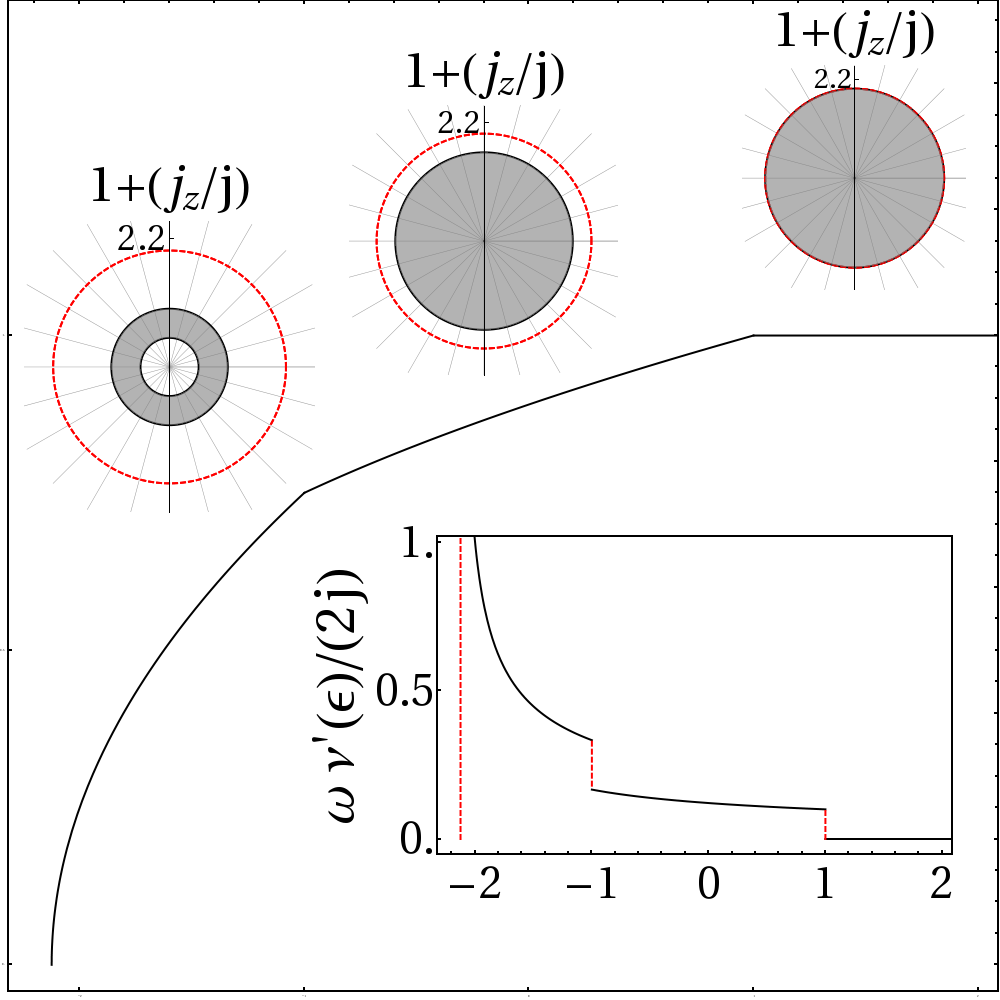}
\\
 \rotatebox{90}{\ \  \ \quad\quad \quad\quad \quad\quad \Large{$\omega \nu(\epsilon)/(2 j)$}}\ \ \ \includegraphics[angle=0,width=.32\textwidth]{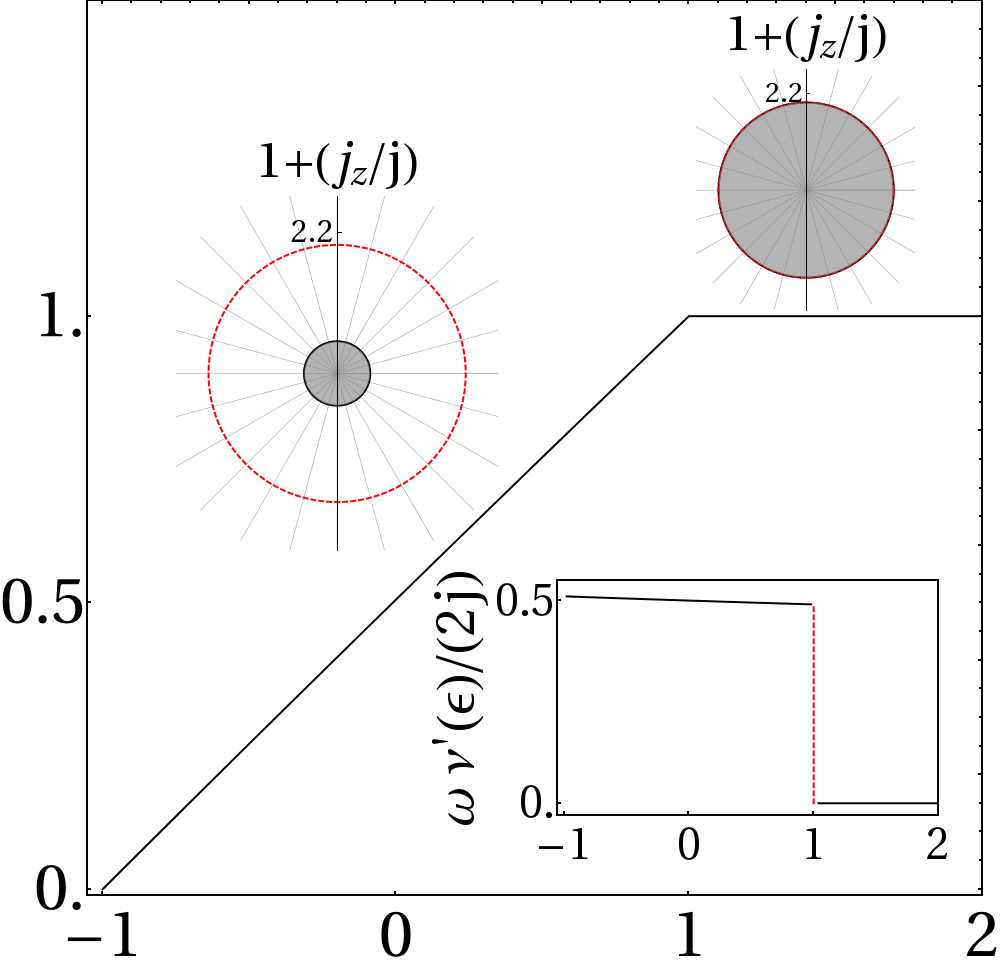}
&
 \includegraphics[angle=0,width=.30\textwidth]{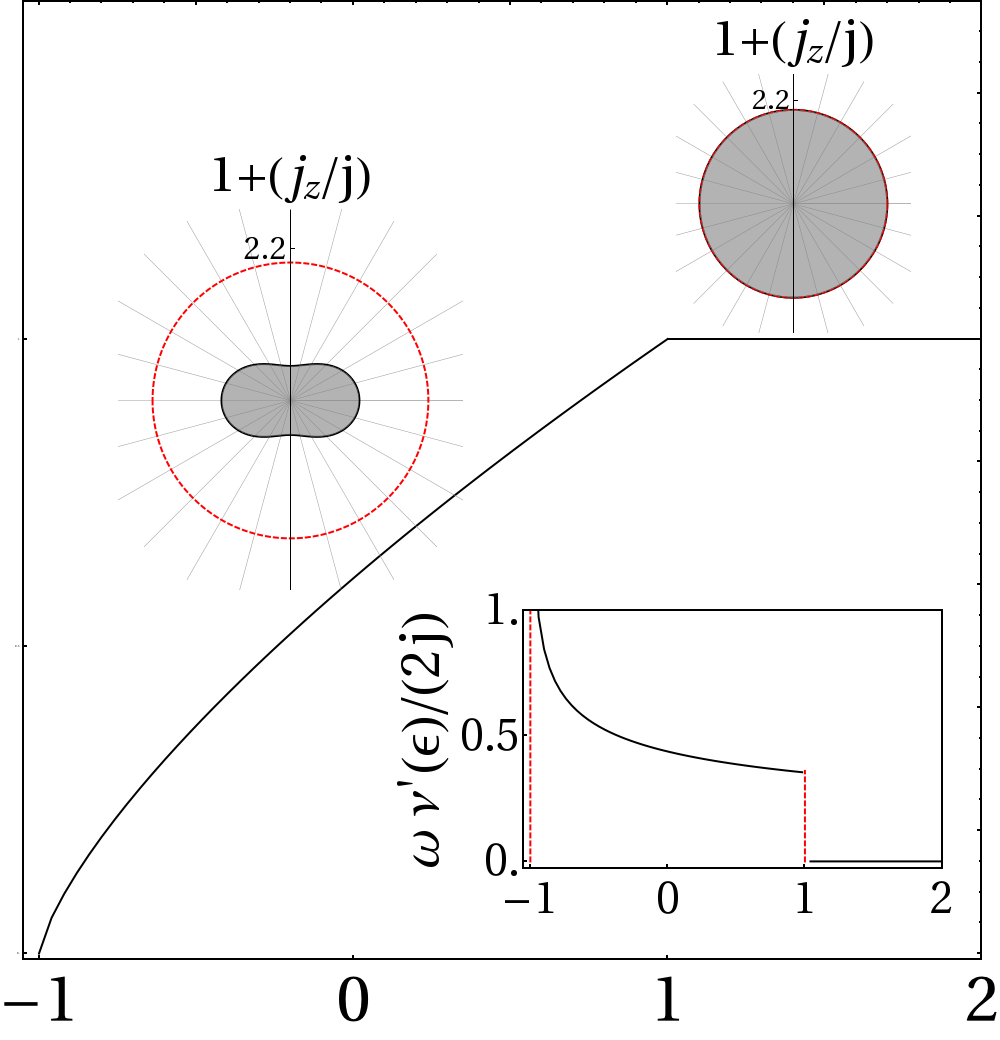}
&
\ \vtop{\vskip -0.313\textwidth \hbox{\includegraphics[angle=0,width=.295\textwidth]{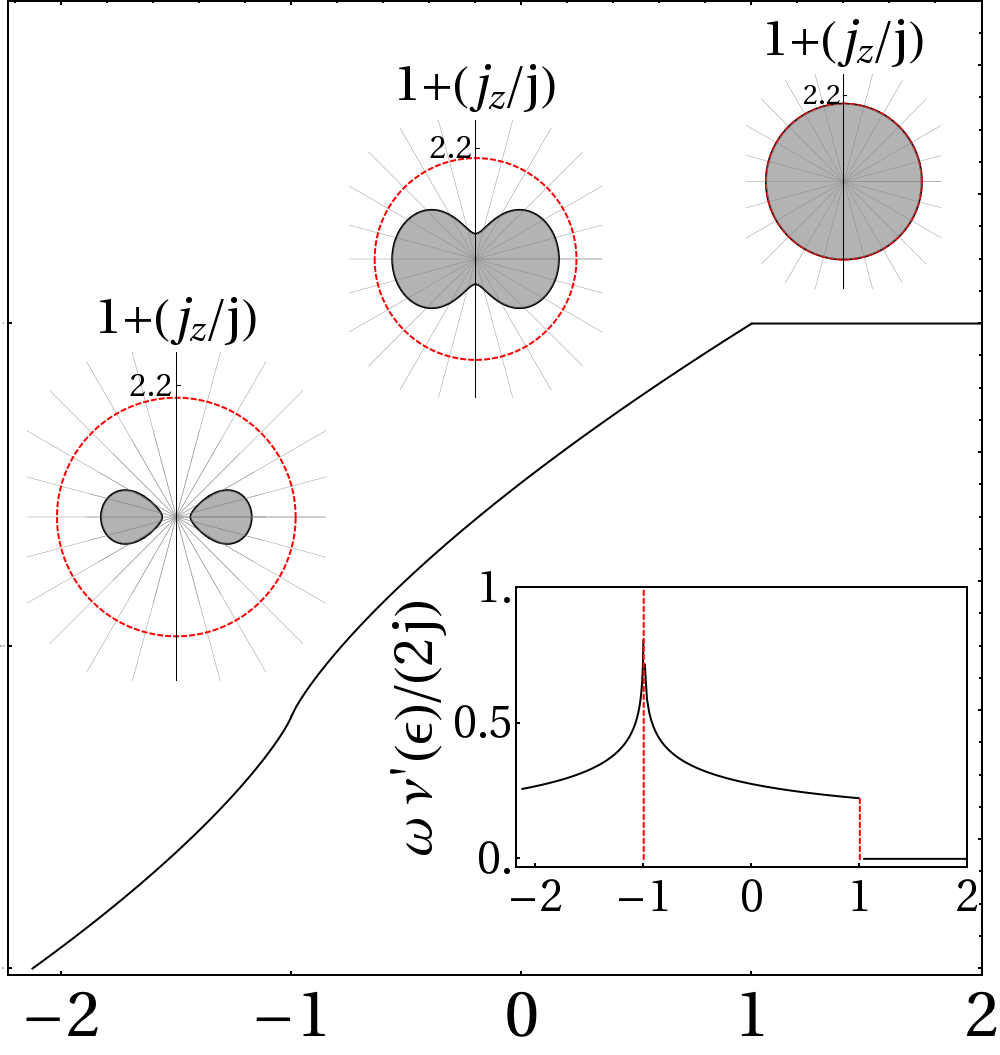}}}
\\
& \LARGE{$\mathbf{\epsilon}$}& 
\end{tabular}}
\caption{ Scaled available phase space volume $\omega \nu(\epsilon)/(2j)$ for the Tavis-Cummings model (top) and Dicke model (bottom) as a function of $\epsilon \equiv E/(\omega_o j)$,  for  couplings $\gamma=0.2 \gamma_c$ (left),  $\gamma=\gamma_c$ (center), and $\gamma=2 \gamma_c$ (right). The derivatives of $\nu(\epsilon)$ are shown as inserts at the bottom right of each panel.
Polar plots, $1+ (j_z/j)$ {\em  vs} $\phi$,  of the available pseudo-spin phase space (gray zones) for  representative energies ( $\epsilon=-0.5$ and $1.5$) are shown in the upper part of the panels. In the two panels on the right a third polar plot is added depicting the available phase space in the superradiant region, for $\epsilon=-2.0$ (TC, top) and $\epsilon=-1.6$ (Dicke, bottom).}
\label{Fig:volD}
\end{figure*}

The volume of the available phase space for a given energy ($E$), which divided by $(2\pi \hbar)^2$ with $\hbar=1$, is given by
\begin{equation}
\nu(E)=\frac{1}{(2 \pi)^2}\int dq \,dp \,d\phi \,d j_z  \,\delta(E-H_{cl}(q,p,\phi,j_z)).
\label{intVol}
\end{equation} 
The previous expression, according to the Gutzwiller's  trace formula \cite{Gutz}, is the semiclassical approximation of the quantum density of states. 
%Recently the previous integral was evaluated  \cite{Bran13} by calculating, first, the partition function of the model, from there the available phase space volume can be extracted as an inverse Laplace transformation. 
Recently this volume was evaluated as an inverse Laplace transform of the partition function of the model \cite{Bran13}. 
Alternatively, 
%it is easy to
we calculate the integral directly. The quadratic nature of the Hamiltonians for the boson variables allows to perform the integrals over $p$ and $q$ giving (see Appendix \ref{app1}) 
\begin{equation}
\nu(E)= \frac{1}{2 \pi \omega}\int d j_z \int d\phi.
\label{intVol2}
\end{equation}
To evaluate this expression we need to know the range of  the pseudospin  variables for a given energy $E$. Here we present the 
main results, the details are shown
in  Appendix \ref{app1}. For the Tavis-Cummings  model the $\Lambda$ symmetry allows the angle variable $\phi$ take any value in the interval $[0,2 \pi)$  for any coupling and energy. Therefore the Eq. (\ref{intVol2}) reduces to   $\nu(E)=(1/\omega)\int dj_z$. On the other hand, the  values the variable $j_z$  can take depend on coupling and energy. Three different energy regimes are identified (a) $1< \epsilon $, (b) -$1\leq\epsilon\leq1$ and (c)  $\epsilon_o\leq \epsilon< -1$, with  $\epsilon_o=-\frac{1}{2}\left(\frac{\gamma_c^ 2}{\gamma^2}+\frac{\gamma^2}{\gamma_c^2}\right)$. The latter  interval appears only in the superradiant phase ($\gamma\geq\gamma_c$).  For energies $1<\epsilon$ the whole pseudo-spin sphere is available: $j_z \in [-j,j]$ and,  consequently, the available  phase space volume saturates ($\nu=2j/\omega$). For energies  $-1\leq \epsilon \leq 1$,   the $j_z$ variable takes values only in the interval $[-j,  j y_+ ]$ with $y_{\pm,}$  ($|y_+|<1$)  given by 
\begin{equation} 
y_{\pm}= \left(-\frac{\gamma_c^2}{\gamma^2} \pm \frac{\gamma_c}{\gamma}\sqrt{2(\epsilon-\epsilon_o)} \right).\label{ypm}
\end{equation}
For couplings above the critical value, $\gamma>\gamma_c $, according to  Eq.(\ref{enesClass}),   the range of possible energies extends until $\epsilon_o<-1$. For the interval $\epsilon\in [\epsilon_o,-1)$  the south pole of the pseudospin sphere ($j_z=-j$)  is inaccessible and the $j_z$ variable is restricted to the interval  $j y_- \leq j \leq j y_+$, with $|y_{\pm}|<1$ given by  Eq.(\ref{ypm}). With the previous results the  classical approximation for the density of states in the Tavis-Cummings model can be easily obtained
\begin{equation}
\frac{\omega}{2 j} \nu(\epsilon)= \left\{ 
\begin{array}{l}
 \frac{\gamma_c}{\gamma}  \sqrt{2(\epsilon - \epsilon_0)}, \ \ \ \ \ \ \ \ \ \ \ \epsilon_0 \leq \epsilon < -1 \\
 \frac {1} {2} \left( 1 - \frac{\gamma_c^2}{\gamma^2} + \frac{\gamma_c}{\gamma} \sqrt{2(\epsilon - \epsilon_0)}\right), |\epsilon|\leq 1 \\
 1,\ \ \ \ \ \ \ \ \ \ \ \ \ \ \ \ \ \ \ \ \ \ \ \ \ \  \ \ \ \ \ \ \ \ \ \ 	\epsilon > 1  .
\end{array}
\right.
\label{dosTC}
\end{equation}

The volume of the  available phase-space  for the Tavis-Cummings model  for three  different couplings, as  a function of the energy, is shown in  the top panels of  Fig.\ref{Fig:volD}. 
The available phase space in the pseudospin-space for different energy regimes is also shown above the curves,
 indicated by gray zones in the polar plots, $1+ (j_z/j)$ {\em  vs} $\phi$. The changes in the available phase space that occurs at energies $\epsilon=\mp 1$,  are clearly indicated by discontinuities in the derivatives $\nu'(\epsilon)$, shown as inserts at the bottom of each panel.
% of  the same figure.     

For the Dicke model the range of the $j_z$ variable  is  (see Appendix \ref{app1}) given by the same expressions as in the Tavis-Cummings model: $j_z\in[j y_-, jy_+]$ for $\epsilon_o~\leq~\epsilon~< ~-~1$, $j_z\in [-j, j y_+]$ for $-1\leq \epsilon \leq 1$, and $j_z\in [-j,j]$ for $1<\epsilon$. On the other hand, since   the $\Lambda$ symmetry is broken for the Dicke model, the available range of the $\phi$ variables depends on coupling and energy. For energies $1<\epsilon$, as in the Tavis-Cummings model, the available pseudo-spin phase space  saturates and $\phi$ takes values in the whole interval $[0,2 \pi)$. For energies $-1\leq \epsilon \leq 1$ the whole interval $[0,2 \pi)$ is accessible only if $-j \leq j_z\leq \epsilon j$. For $j \epsilon <j_z \leq j y_+$ the $\phi$ variable is restricted by the condition 
\begin{equation} 2j\frac{\gamma_c^2}{\gamma^2}\frac{(j_z-j\epsilon)}{(j^2-j_z^2)}\leq \cos^2\phi\leq 1 .\label{phiCond}
\end{equation} 
Finally for $\epsilon_o\leq \epsilon<-1$ (possible only in the superradiant phase $\gamma>\gamma_c$), the $\phi$ variable is restricted by the same  condition (\ref{phiCond}). Having identified the range of the pseudospin variable, it is straightforward to obtain the following expression  for $\nu(\epsilon)$ 
for the Dicke model
 \begin{equation}
\frac{\omega}{2 j} \nu(\epsilon)= \left\{ 
\begin{array}{l}
\frac{1}{\pi}\int_{y_-}^{y_+} \arccos \sqrt{\frac{2\gamma_c^2 (y-\epsilon)}{\gamma^2(1-y^2)}} dy,   \ \ \ \ \epsilon_0 \leq \epsilon < -1 \\
\frac{\epsilon+1}{2}+\frac{1}{\pi}\int_{\epsilon}^{y_+} \arccos \sqrt{\frac{2\gamma_c^2 (y-\epsilon)}{\gamma^2(1-y^2)}} dy  , \ \ \ |\epsilon|\leq 1 \\
 1,\ \ \ \ \ \ \ \ \ \ \ \ \ \ \ \ \ \ \ \ \ \ \ \ \ \  \ \ \ \ \ \ \  \ \ \  \ \ \ \ \ \ \  \ \ 	\epsilon > 1 ,
\end{array}
\right.
\label{dosDicke}
\end{equation} \ 
where $y_\pm$ is given by (\ref{ypm}). 

The previous expression for the available phase space volume is plotted in Fig. \ref{Fig:volD} for three couplings as a function of the energy, 
in the lower panels. 
The available pseudospin phase space  for energies in the different regimes is also shown above the curves as gray areas in the polar plots. The changes in the available phase space occurring at energies  $\epsilon=\mp 1$ are 
% clearly indicated by 
evident as
discontinuities and divergences in the derivative $\nu'(\epsilon)$. Observe that for small couplings ($\gamma=0.2\gamma_c$, left)  the Dicke and Tavis-Cummings curves are very similar, but they differ clearly at  the critical coupling, where  the available regions in the Dicke model are highly deformed.  The differences are more dramatic  in the superradiant phase   $\gamma>\gamma_c$: while a discontinuity in the first derivative occurs at $\epsilon=-1$ for the TC model, the non-analytic  behavior of the derivative  of $\nu(\epsilon)$ in the  Dicke model  is a logarithmic  divergence  \cite{Bran13}. This behavior can be understood by looking at the geometry of the available phase space in both models. 
%while i
In the TC model  the available phase consists of a single circularly symmetric connected region,
but  in the Dicke model it consists of  two disconnected regions for $\epsilon<-1$, which touch each other in the saddle point at $\epsilon=-1$ and  merge for larger energies $\epsilon>-1$. In the next two subsections, the previous classical approximations for the Density of States are compared with the results coming from diagonalizing the Hamiltonian of the Tavis-Cummings and Dicke quantum models.  

%%%%%%%%%%%%%%%%

\subsection{Quantum density of states in the Tavis-Cummings model}

\begin{figure}
\centering{
\begin{tabular}{cc}
(a) $ \gamma=\gamma_{c}$ & (b) $ \gamma= 2 \,\gamma_{c}$ \\ 
\ \ \rotatebox{90}{\  \ \    \quad\quad \quad \large{$n/j$}} \includegraphics[angle=0,width=0.2\textwidth]{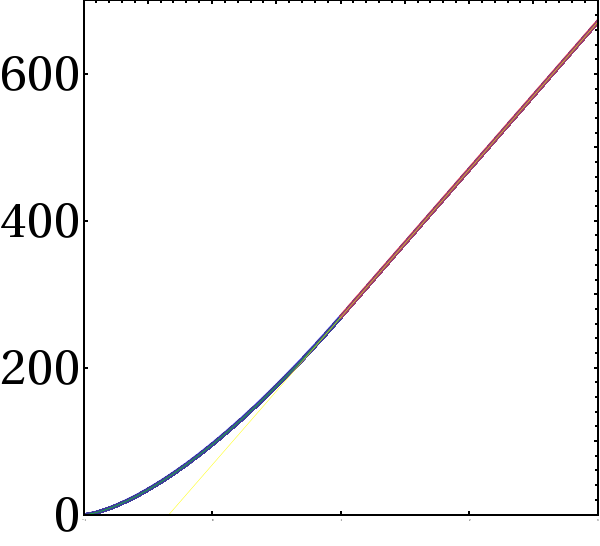}%{TCj100_n_vs_e_g1v2.png}
& 
 \includegraphics[angle=0,width=0.2\textwidth]{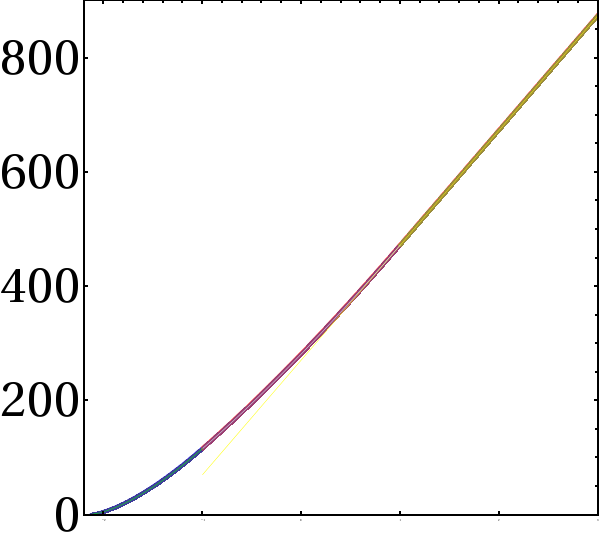}%{TCj100_n_vs_e_g2v2.png}
 \\
 
\qquad \ \Large{$\mathbf{\epsilon}$}&\qquad \ \Large{$\mathbf{\epsilon}$}\\
\rotatebox{90}{\  \ \     \quad \large{$(\omega/2j) \Delta \bar{n}/\Delta \bar{E}$}} \ \ \  \includegraphics[angle=0,width=0.2\textwidth]{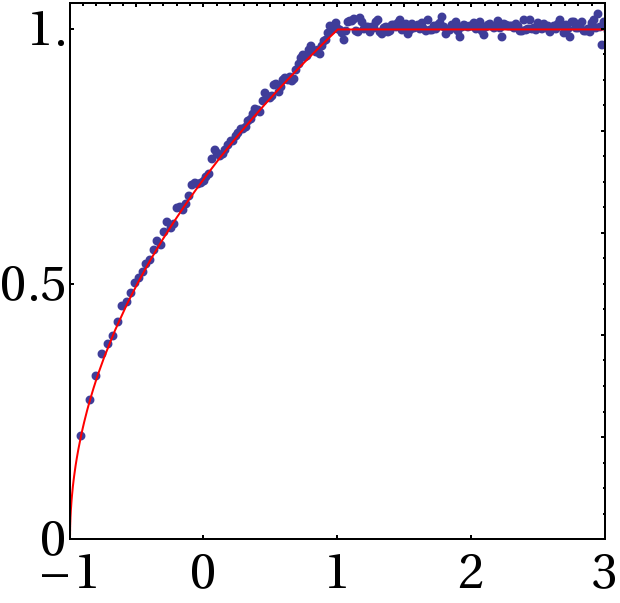}%{DenTCf1.pdf}
 &
\ \  \includegraphics[angle=0,width=0.2\textwidth]{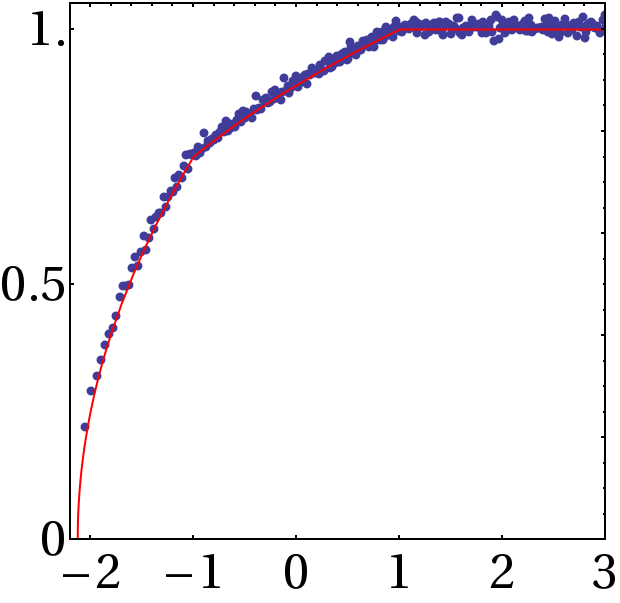}%{DenTCf2.pdf}
\\
\qquad \ \Large{$\mathbf{\bar{\epsilon}}$}&\qquad \ \Large{$\mathbf{\bar{\epsilon}}$}
\end{tabular}
}
\caption{(Color online) Top row: $\frac n j$ as function of $\epsilon$ for (a) $ \gamma=\gamma_{c}$ and (b) $ \gamma= 2 \, \gamma_{c}$. Bottom row: Averaged quantum density of states, $\frac{\omega}{2j} \frac {\Delta \bar n}{\Delta \bar E}$ (blue points), in the Tavis-Cummings model  as a function of $\bar \epsilon$,  for (a) $\gamma=\gamma_{c}$ and  (b) $ \gamma= 2 \, \gamma_{c}$. The continuous red  lines depict the semi-classical results.}
\label{fig4}
\end{figure}

The basis in which the Tavis-Cummings Hamiltonian is diagonalized, for fixed $j$, can be labeled by $\lambda$ and $m$. 
For a given value of $\lambda$, the number of states in each subspace is N$_{st}(\lambda)$ = Min($\lambda+1$, $2j+1$). 
%The 
This
number of states grows linearly with $\lambda$ up to $\lambda_0 = 2j$, and from that value on it remains fixed in $2j +1$. It represents a {\em static} change in the density of states which is always present.
The eigenstates of $H_{TC}$ can be classified as $E(i,\lambda)$, $i=1$, N$_{st}(\lambda)$.
To obtain a complete energy spectrum up to an energy $E_{ref}$, all subspaces up to $\lambda_{max}$ must be included, where Min($E(i,\lambda_{max})$)$>$ $E_{ref}$.

We have studied the resonant case, $\omega =\omega_o = 1$, which has $\gamma_{c,TC}=1.0$. Selecting $\mathcal{N}=200$ ($j=100$), $\lambda_{max}=2000$ is enough to provide the complete energy spectrum up to the scaled energy $\epsilon = 6.4$ for $ \gamma=\gamma_{c,TC}$, with 264000 states, and up to $\epsilon = 3.3$ for $ \gamma= 2 \, \gamma_{c,TC}$ with 160,000 states. 

Using $n$ as the order number in which each state with energy $E$
%%%$\epsilon$
 appears in the energy spectrum, in top row of Fig.\ref{fig4} we present $\frac{n}{\mathcal{N}}$ as function of $\epsilon$ for (a) $ \gamma=\gamma_{c,TC}$, (b) $ \gamma= 2 \, \gamma_{c,TC}$. 
Notice the three regions, displayed with different colors, corresponding to $\epsilon_0 \leq \epsilon < -1$, $ |\epsilon|\leq 1$
and $\epsilon > 1$ in Fig. \ref{fig4}(b), while in Fig.\ref{fig4}(a) there are only two, because  the ground state energy is $\epsilon_{GS}=E_{GS}/(\omega_o j)= -1$. The thin lines inside 
are the fits in each region, inspired in their functional form in the integrals of $\nu(\epsilon)$. 
It is worth to mention that the derivatives of the fitted function coincide with $\nu(\epsilon)$, with differences of the order $\frac{1}{\mathcal{N}}$.

While the curves presented in the top row of Fig.\ref{fig4} seem to be smooth, obtaining their first derivative as finite differences in order to estimate the quantum the states,  is tricky because the fluctuations obscure the results. To overcome this difficulty we have taken averages of the energy
%%% $\bar\epsilon(\bar n)$
 $\bar E(\bar n)$ over intervals of 600 levels, with average number of state $\bar n$. From these averaged quantities we obtain the average derivative
%%% $\frac {\Delta \bar n}{\Delta \bar\epsilon}$,
$\frac {\Delta \bar n}{\Delta \bar E}$,  displayed in the bottom row of Fig.\ref{fig4}.

The 
%%%thin green curves
continuous red curves representing $\nu(\epsilon)$ overlap nicely with the averaged numerical results, presented as points. The {\em static} excited-state phase transition at $\epsilon=1$ is present in both cases, while the \emph{dynamic} phase transition at $\epsilon= -1$ can be observed, very clearly, for the super radiant case 
$ \gamma= 2 \, \gamma_{c,TC}$.

\subsection{Quantum density of states in the Dicke model}

We repeat some of the calculations we did in the case of the TC model for the Dicke model, but in this case we must be careful with the convergence of the numerical solutions as the model is not integrable. We diagonalize numerically the Dicke Hamiltonian employing an extended bosonic coherent basis (see Appendix \ref{app2}), which let us obtain a significative part of the energy spectra with a small truncation or cutoff \cite{Chen0809,Basta11,Basta12}. For a given truncation we can estimate for each individual excited state a lower bound of the numerical precision in the wave function, as it is pointed out in Appendix \ref{app2}.  
In this way we can monitor that each eigenstate has converged 
up to some chosen significative figures. We have selected the resonant case $\omega=\omega_o$, with $\mathcal{N}=80$ ($j=40$).

\begin{figure}
\centering{
\begin{tabular}{c}
%$\gamma= \gamma_{c}$\\
\rotatebox{90}{\quad  \quad     \quad \Large{$(\omega/2j) \Delta \bar{n}/\Delta \bar{E}$}} \includegraphics[angle=0,width=0.43\textwidth]{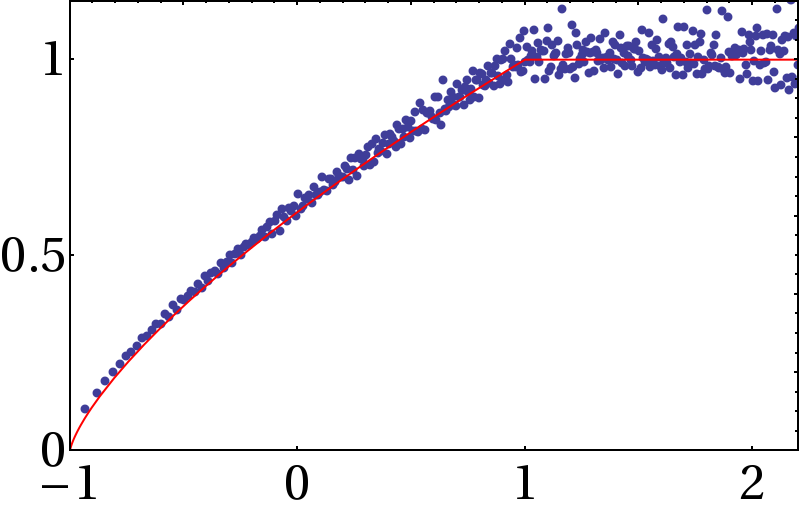}%{nexc_av_vs_e_g0_5v2.png}
\\ 
%$\gamma= 2\,\gamma_{c}$\\ 
\rotatebox{90}{\quad  \quad     \quad \Large{$(\omega/2j) \Delta \bar{n}/\Delta \bar{E}$}} \includegraphics[angle=0,width=0.43\textwidth] {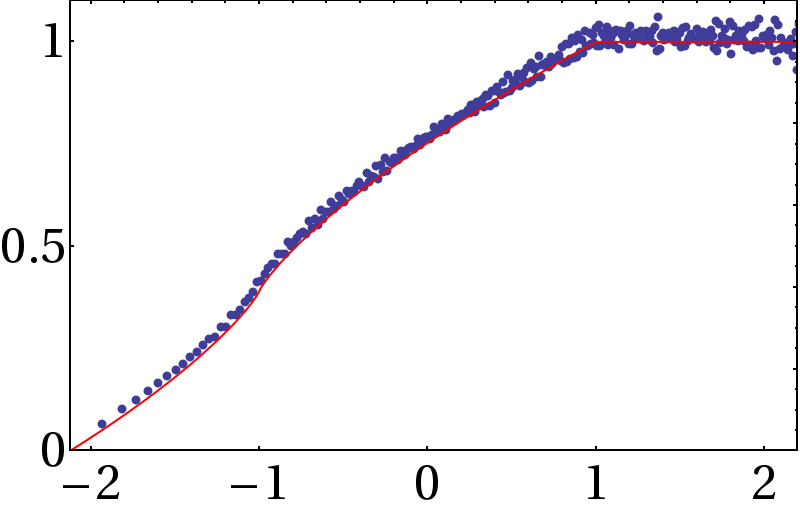}%{nexc_av_vs_e_g1_0.png}
\\
\qquad\qquad\LARGE{$\mathbf{\bar{\epsilon}}$}
\end{tabular}
}
\caption{(Color online)
%%%$\frac {\Delta \bar n}{\Delta \bar \epsilon}$, 
Averaged quantum density of sates, $\frac{\omega}{2j}\frac {\Delta \bar n}{\Delta \bar E}$ (blue points), in the Dicke model   as a function of $\bar \epsilon$ for  $ \gamma=\gamma_{c}$ (top) and $ \gamma= 2 \, \gamma_{c}$ (bottom). The continuous red lines indicate  the corresponding semi-classical results.}
\label{fig5}
\end{figure}

For the Dicke model the fluctuations in energy are smaller than in the TC, and the averages of the energy 
%%%$\bar\epsilon(\bar n)$ 
$\bar E(\bar n)$ 
are taken over intervals of 
20  levels, with average number of state $\bar n$. From these averaged quantities we obtain the average derivative
%%% $\frac {\Delta \bar n}{\Delta \bar\epsilon}$
$\frac {\Delta \bar n}{\Delta \bar E}$, displayed in Fig.  \ref{fig5}
 
The 
%%%thin green
continuous red  curves plot $\nu(\epsilon)$, the same ones plotted in Fig. \ref{Fig:volD}, which also in this case overlap nicely with the averaged numerical results, presented as points. The {\em static} ESQPT at $\epsilon=1$ is present in both cases, while the {\em dynamic} phase transition at $\epsilon= -1$ can be observed, very clearly, for the super radiant case
%%% $ \gamma= 3 \, \gamma_{c}$.
$ \gamma= 2 \, \gamma_{c}$.

The numerical evidence provided in this section shows that the semi-classical density of states 
%calculated in the previous section 
describes correctly the  tendency of the quantum spectra of the Tavis-Cummings and Dicke model, both in the normal and super-radiant phases. Consequently, the semi-classical result can be safely used  to perform the so-called unfolding of the quantum spectra and  study the statistical properties of quantum fluctuations. It is 
%very 
well  known that the properties of these fluctuations are 
%common with 
the same as
those of different random matrix ensembles depending on the dynamic of the underlying semi-classical model: the gaussian diagonal ensemble (GDE) for quasi-integrable or regular dynamics, and the gaussian orthogonal ensemble (GOE)  for chaotic dynamics with time invariant symmetry. 
%The former 
This
analysis is performed  in the companion paper \cite{Basta2} to this one.    

%%%%%%%%%%%%%%%%

\section{Conclusions}

Using both a semi-classical analysis and  results of an efficient numerical procedure to diagonalize the quantum Hamiltonians,  we have studied  the Dicke and Tavis-Cumming models in the space of couplings and excitation energies.
We have focused on a global property in the energy-coupling space: the excited-state quantum phase transitions or singular behavior of the density of states.

 Analytical results for the semi-classical approximation to the density of sates were derived by calculating the volume of the available phase space for a given coupling and  energy.  From the classical analysis, two different unstable fixed points of the Hamiltonian flux can be identified. The first one located at the north pole of the pseudo-spin sphere appears for any coupling. The second one appears only in the superradiant phase and is located at the south pole of the pseudo-spin sphere. The role of these unstable fixed points in relation to  the occurrence of the excited-state quantum phase transitions (ESQPTs) was discussed and established. The unstable points are benchmarks in   the energy space which indicate an abrupt change in the available phase space. The two unstable fixed points produce two kinds of ESQPTs. The first one,  referred to  as {\em static} ESQPT, occurs for any coupling at energy $E/(\omega_o j)=1$.  At this energy the whole pseudo-spin sphere becomes available for the system.  The second ESQPT, referred to as {\em dynamic}, occurs only  for couplings larger than the critical one  at energies $E/(\omega_o j)=-1$. This transition occurs when the top of the double well (Dicke) or mexican hat (TC)  potential  that develops in the superradiant phase is attained.  The abrupt changes in the available phase space are reflected  in the classical density of states as non-analytic behavior of its first  derivative. For the integrable TC  model, the first derivative shows a discontinuity for both the {\em static} and {\em dynamic} ESQPTs. For the Dicke model the {\em static} ESQPT is equally reflected by a discontinuity of the first derivative, but the {\em dynamic} ESQPT is associated with a logarithmic divergence of the first derivative. For the quantum case, finite systems [$\mathcal{N}=200$  (TC) and $\mathcal{N}=80$  (Dicke)] were diagonalized in large energy regions which include all the regimes identified in the semi-classical approximation. The tendency of the quantum spectra was obtained by averaging the energy and the number of state index, over intervals of $600$ (TC) and
$20$ (Dicke) contiguous  states. After this average procedure, it was shown that  the quantum results overlap perfectly with the semi-classical density of states. This result confirms that the semi-classical approximation is appropriate to perform the unfolding of the quantum spectrum, and consequently to study the properties of its fluctuations \cite{Basta2}.

We thank P. Str\'ansky and P. Cejnar for many useful and interesting conversations.This work was partially supported by CONACyT- M\'exico,  DGAPA-UNAM and DGDAEIA-UV through the ''2013 Internal call for  strengthening  academic groups" (UV-CA-320).

%%%%%%%%%%%%%%%%

\appendix

\section{Available phase space for a given $E$}
\label{app1}
Here, we perform the boson variables ($q$ and $p$) integration of [with $H_{cl}(q,p,\phi,j_z)$ defined in Eq.(\ref{DiHam})]
$$
\nu(E)=\frac{1}{(2 \pi)^2}\int d j_z  \,d\phi \, dp \, dq  \, \delta(E-H_{cl}(q,p,\phi,j_z)),
$$
and determine the range of the pseudospin variables for a given energy and coupling. The $q$ integration is straightforward by using the properties of the Dirac delta,
$$
  \nu(E)=\frac{1}{(2 \pi)^2}\int d j_z \, d\phi \, dp \, dq \, \left( \frac{\delta (q-q_+)}{|\partial H/\partial q|_{q_+}}+ \frac{\delta (q-q_-)}{|\partial H/\partial q|_{q_-}} \right), 
$$
where $q_\pm$ are the roots of the quadratic equation $E~-~H(q,p,\phi,j_z)=0$: 
\begin{equation}
\omega q_{\pm}=-\gamma\sqrt{j}\cos\phi\sqrt{1-\frac{j_z^2}{j^2}} (1+\delta)\pm  \sqrt{-\omega^2 p^2+b \,p+c},
\label{qfunction} 
\end{equation}
with  the coefficients $b$ and $c$  given by $$b=2\omega\gamma\sqrt{j}\sin\phi\sqrt{1-\frac{j_z^2}{j^2}} (1-\delta)$$ and $$c=\gamma^2 j\cos^2\phi\left(1-\frac{j_z^2}{j^2}\right)(1+\delta)^2+2\omega(E-\omega_o j_z).$$ Evaluating the derivatives, one obtains  $|\partial H/\partial q|_{q_+}= |\partial H/\partial q|_{q_-}=\sqrt{-\omega^2 p^2+b p+c}$, then the $q$
integration yields
$$
 \nu(E)=\frac{1}{(2 \pi)^2}\int d j_z \, d\phi \, dp  \, \frac{2}{\sqrt{-\omega^2 p^2+b p+c}}, 
$$
with the limits in the variables $j_z$, $\phi$, and $p$ determined by the condition $-\omega^2 p^2+b p+c\geq 0$. The $p$ integration is easily performed by writing $$-\omega^2 p^2+b p+c= \omega^2 (p_+-p)(p-p_-),$$ with $p_{\pm}$ the roots ($p_- \leq p_+$) of the quadratic polynomial  $-\omega^2 p^2+b p+c=0$, 
\begin{eqnarray}
 \nu(E)&=&\frac{2}{\omega (2 \pi)^2}\int d j_z \int  d\phi \int_{p_-}^{p_+}dp  \frac{1}{\sqrt{(p_+-p)(p-p_-)}}\nonumber\\
       &=& \frac{2 \pi }{\omega (2 \pi)^2}\int d j_z \int  d\phi ,\nonumber 
\end{eqnarray}
The previous result is valid provided that the roots $p_{\pm}$ are real, which in turn occurs only if the maximum of the polynomial $-\omega^2 p^2+b p+c$ is greater or equal than zero:
$$
\frac{b^2}{4\omega^2}+c\geq 0.
$$
By substituting the values of $b$ and $c$, the previous condition reads
\begin{equation}
\frac{\gamma^2}{2 \gamma_c^2}\left(1-y^2 \right)\left (\frac{(1-\delta)^2}{(1+\delta)^2}\sin^2\phi+\cos^2\phi)\right)\geq y-\epsilon,
\label{conda}
\end{equation}
with  $\gamma_c=\sqrt{\omega\omega_o}/(1+\delta)$, and  we have used  the variables $y\equiv j_z/j$ ($|y|\leq 1$) and $\epsilon\equiv E/(\omega_o j)$. The previous condition determines the range of the pseudospin variables for a given energy $\epsilon$. For the Tavis-Cummings model ($\delta=0$) the previous condition is independent of $\phi$ and  simplifies to 
$$
\frac{\gamma^2}{2 \gamma_c^2}\left(1-y^2 \right)\geq y-\epsilon,
$$ 
therefore no  restriction for  the variable $\phi$ occurs and it can take any value in the interval $[0,2\pi)$. If $\epsilon>1$ the previous condition is satisfied  in the whole   interval $y\in[-1,1]$, therefore in this case the whole pseudospin sphere is accessible. For $-1\leq\epsilon\leq 1$, the condition is satisfied only for $y\in[-1,y_+]$ ($y_+<1$) where    $y_{\pm}$   are the roots of $\frac{\gamma^2}{2 \gamma_c^2}\left(1-y^2 \right)= y-\epsilon$, given in Eq.(\ref{ypm}). Finally, for energies $\epsilon<-1$, the condition is  satisfied in the interval $y\in[y_-,y_+]$ ($|y_{\pm}|<1$) only if  $\gamma>\gamma_c$ and $\epsilon\geq\epsilon_o$, where $\epsilon_o<-1$ is the classical ground-state  energy  in the superradiant phase defined immediately after the Eq.(\ref{ypm}).  

For the Dicke model ($\delta=1$) the condition  (\ref{conda}) is
\begin{equation}
\frac{\gamma_c^2}{\gamma^2}\frac{2(y-\epsilon)}{1-y^2}\leq \cos^2\phi,
\label{condphi}
\end{equation}
clearly, this condition constrains the values the $\phi$ variable can take. If $\epsilon>1$ the condition is satisfied for the whole  pseudospin sphere  $y\in[-1,1]$ ($j_z\in[-j,j]$) and $\phi\in[0,2\pi)$. For energies satisfying $-1\leq\epsilon\leq 1$, similar to  the Tavis-Cummings case, the condition can be satisfied only for $y\in[-1,y_+]$, but, here, contrary to the Tavis-Cummings case, a restriction to the $\phi$ variable appears as follows: if $y\in[-1,\epsilon]$,  $\phi$ takes values in the whole interval $[0,2\pi)$, but if 
  $\epsilon<y\leq y_+$ the angular variable is restricted by the condition  (\ref{condphi}),
which is satisfied for values in  intervals around $\phi=0$ and $\phi=\pi$. 

Finally, as in the Tavis-Cummings case, for energies $\epsilon<-1$, the condition can be satisfied in the interval $y\in[y_-,y_+]$ only if  $\gamma>\gamma_c$ and $\epsilon\geq\epsilon_o$, where $\epsilon_o<-1$ is the classical ground-state energy in the superradiant phase. But now, contrary to the Tavis-Cummings case, the angular variable is restricted by the condition (\ref{condphi}).  

%%%%%%%%%%%%%%%%

\section{Numerical solutions and precision in the wave function}
\label{app2}
We use an extended bosonic coherent basis in order to diagonalize the Dicke Hamiltonian  \cite{Chen0809,Basta11,Basta12}. The basis corresponds to the eigenstates of the  Dicke model's integrable limit $\omega_{0}\rightarrow 0$. 
We write it as $|N;j,m'\rangle$, where $m'$ are the eigenvalues of $J_{x}$ and $N$ is the eigenvalue of the $A^{\dagger}A$ operator, with $
A=a+\frac{2\gamma}{\sqrt{\mathcal{N}}\omega}J_{x}$,
\begin{equation}
|N;j,m'\rangle=\frac{1}{\sqrt{N!}}(A^\dagger)^N |N=0; j,m'\rangle.
\end{equation}
The vacuum for a given $m'$ is a boson coherent state ($|\alpha\rangle$) times an eigenstate of the $J_x$ operator: $$   | N=0; j,m'\rangle=\left|\alpha=-\frac{2 \gamma m'}{\omega_o \sqrt{\mathcal{N}}}\right\rangle |j m'\rangle.$$  
Now, the $k$th excited state wave function of the Dicke Hamiltonian can be written as
\begin{equation}
|\Psi^{k}(N_{max})\rangle=\sum\limits_{N=0}^{N_{max}} \sum\limits_{m'=-j}^{j} C^{k}_{N,m'} |N;j,m'\rangle.
\end{equation} 
Here, $C^{k}_{N,m'}$ are the coefficients of the  $k$th wave function in terms of the extended bosonic coherent basis and $N_{max}$ is the value of the truncation or cutoff in the number of displaced excitations ($0\leq N\leq N_{max}$). The probability $P_{N}$ of having $N$ excitations in the $k$th state is,
\begin{equation}
P^{k}_{N}=|\langle N|\Psi^{k}\rangle|^{2}=\sum_{m'}|C^{k}_{N,m'}|^{2}.
\end{equation}
We define the precision in the calculated wave function as \cite{Basta13}:
\begin{equation}
\Delta P^{k}=\sum\limits_{m'=-j}^j \left|C^{k}_{N_{max}+1,m'}\right|^2.
\end{equation}
By diagonalizing the Hamiltonian with several truncations, we consider that the solution has converged if $ \Delta P^{k}$ is smaller than certain tolerance, being $N_{max}$ the minimum value of the truncation necessary for obtaining the  numerical solution to the desired precision. 
 
%%%%%%%%%%%%%%%%

%%%%%%%%%%%%%%%%%%%%%%%%%%%%%%%%%%%%%%%%%%%%%

%%%%%%%%%%%%%%%%
\end{document}